\documentclass[epj,referee]{svjour}

\usepackage{amssymb}
\usepackage{epsfig}

\begin{document}
\title{Electron interference and entanglement in coupled 1D systems with noise}

\author{Fabrizio~Buscemi\inst{1} \and Paolo~Bordone\inst{2,3} \and Andrea Bertoni\inst{3}}

\institute{ARCES, Alma Mater Studiorum, Universit\`{a} di Bologna, Via Toffano 2/2, 40125 Bologna, Italy, \email{fabrizio.buscemi@unimore.it}
\and Dipartimento di Scienze Fisiche, Informatiche e Matematiche, Universit\`{a} di Modena e Reggio Emilia, Via Campi 213/A, 41125 Modena, Italy
\and Centro S3, CNR-Istituto Nanoscienze, Via Campi 213/A,  41125 Modena, Italy}  

\abstract{We  estimate  the role of   noise in the formation of entanglement and  in the appearance of 
 single- and two-electron interference  in systems  of  coupled  one-dimensional  channels. 
Two cases are considered: a single-particle interferometer and a two-particle interferometer
exploiting  Coulomb interaction. In both of them, environmental noise  yields a  randomization of the carrier phases.
 Our results  assess how  the complementarity
relation linking single-particle behavior  to nonlocal quantities (such as  entanglement and environment-induced decoherence)
acts in electron interferometry.  We show that  in an  experimental implementation of the setups
examined, one- and two-electron detection probability at the output drains can  be used to evaluate the decoherence and  the
degree of entanglement.}

\date{Received: date / Revised version: date}

\maketitle

\section{Introduction}
In the last decades, much attention has been devoted to semiconductor systems 
able to control and expose the correlated dynamics of few carriers~\cite{Grabert,Zahid}.  In particular, the study 
of the electron propagation 
in one-dimensional (1D) structures opened interesting opportunities for
device applications in quantum electronics and optoelectronics, such
as the single-electron transistor~\cite{Kast}. Beyond the design of nanometric
devices, the  coherent charge transport in 1D channels 
 is also very
appealing from the basic physics perspective, as it enables to
analyze quantum  features, such as coherence and interference,
and to relate them to the appearance of quantum correlations~\cite{Dressel,Buscemi2}.
Furthermore, it  also represents a suitable means to fully exploit the
 potential of the entanglement as the basic ingredient of  quantum information processing.
 

Recent experiments showed the feasibility of  electron
interferometers in 1D semiconductor systems, such as quantum wires~\cite{Tilke,Tarucha}
and edge states operating in the integer quantum Hall regime~\cite{Hei,Neder}.  The former,
relying on the quantization of conductance in narrow constrictions
within two-dimensional electron gases, have been the subject of a
number of theoretical~\cite{Akg,Rei,Yang,Eugster} and
experimental~\cite{Tilke,mori,Bird,Bird2} works aimed at  investigating
 their potential applications as  quantum logic devices.  In particular,
the possibility of realizing a universal set of quantum  gates by means  of coupled quantum wires 
was proven in Ref.~\cite{Bertoni}, where the localization of the electron in one
of the two 1D channels  defines a flying qubit. Recently, quantum-wire
systems have been shown to be suitable to implement complex quantum computation
protocols, such as quantum teleportation and Shor's factorization
algorithm~\cite{bus,bus2}.

Most of  the electron devices applications  using 1D channels rely on the coherent propagation
of carriers. As it is universally recognized, the loss of coherence of a quantum system
due to the unavoidable interaction with its environment represents the major threat to the quantumness
of the system itself~\cite{giulini,Zurek}. Also, the working of interferometer setups 
or devices implementing quantum information protocols results to be jeopardized by  decoherence processes.
This justifies the growing attention  recently devoted  to the theoretical 
investigation of mechanisms responsible for the loss of coherence  in quantum interference setups~\cite{Ady,Langen,Davies,Marq,Marq2,Jacobs}. 
 In these systems, the loss of coherence,  resulting in a decrease of the ÒvisibilityÓ of the interference pattern,
has been attributed to the thermal averaging related to the spread of the wavelengths of  the carriers contributing to
the current or to other dephasing mechanisms more intrinsically related to  the environment features. Specifically,
most of the works adopt the phenomenological   ``dephasing terminal''  approach where an additional
artificial electron reservoir randomizes the phase of  electrons~\cite{Langen,Davies}.
In other  schemes, the environment has been modeled as a classical noise
field that leads to a fluctuation of the phase difference for the electrons
crossing the interferometer devices~\cite{Marq,Marq2}.

In this paper, we present a theoretical approach for analyzing the effect 
of  the electron decoherence  on single-electron and two-electron quantum interference and on
carrier-carrier entanglement.  Specifically,
two different systems are examined: a single-electron  Mach-Zehnder interferometer (MZI)
and a device consisting of two  MZIÕs coupled together by the Coulomb interaction operating
in a two-particle regime. We also track the effects of the noise. The two-particle interferometer  has  been analyzed in  a recent work~\cite{Dressel},
where the \emph{which-path} measurement  on one of the two MZIÕs
has  been widely discussed by using the other one as a detector
and neglecting decoherence phenomena due to coupling with the environment.

In our approach, we describe  the environmental noise in terms of randomization of the phases of the charge carriers traveling along the channels. In agreement with the   ``classical fluctuating potential''  procedure~\cite{Ady,Marq}, such a process is ascribed to stochastic fluctuations affecting the  potential profile of the electron waveguides responsible  of phase delay of the electrons.
As a result,  the latter phase obeys a statistical distribution related to the  intrinsic features of the noise.
The quantum system is thus described by the statistical mixture of the states obtained by averaging over the carrier phases.  
As expected, the  quantumness  of the system is reduced by an increasing disorder.
 Such a noise model  represents a valuable guideline to evaluate analytically how the electron dephasing affects interference and entanglement in  mesoscopic setups exploiting electron-electron interaction. In order to validate our approach, first we apply it  to a simple system, that is a one-electron MZI and then we  extend it to a two-electron interferometer. We  focus on the complementarity relations  between single-particle properties, such as single-particle visibility, predictability, and non-local quantities, such as  entanglement with the remaining part of  the system~\cite{Jag1,Jag2,Peng2}. Such  relations, initially  introduced for  pure two-qubit states~\cite{Jag1,Jag2} and then extended to  pure multi-qubit states~\cite{Peng2}, have been validated   both in quantum optics~\cite{Jab2,Suz}  and NMR experiments~\cite{Peng}. 
Our aim is to show that the complementarity relations hold for two-electron mixed states
obtained in noisy 1D technologically  relevant semiconductor channels.

The paper is organized as follows: In Sec.~\ref{SPI}, we illustrate a single-electron  interferometer, implemented by means of 1D structures, and then we analyze the effects of  noise on the visibility of the fringes of the single-particle interference.
In Sec.~\ref{TPI}, we discuss the coupling via Coulomb interaction of two single-electron interferometers in absence of noise and then we study the effect of their entanglement on interference phenomena.
In  Sec.~\ref{TPI2}, we examine how the dephasing due to noise affects both the two-particle interference and the particle-particle entanglement in the two-electron interferometer.
Finally, we present our conclusions in  Sec.~\ref{Conclu}.

\section{Single-particle interferometer}
\label{SPI}
\begin{figure}[h]
  \begin{center}
    \includegraphics*[width=\linewidth]{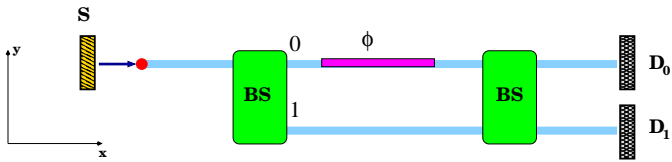}
    \caption{\label{fig1} Sketch of the electron MZI. A charge carrier
      injected from the source $S$ reaches a beam splitter where it is
      split in two components: one propagating along the
      waveguide 0 where  a potential barrier is placed, the other
      traveling in the waveguide 1. Both components merge at the
      second beam splitter, where they interfere.  The particle
      is revealed either at the drain $D_0$ or at the drain $D_1$, e.g.  by means of
      single-electron transistors acting as sensitive charge
      detectors. In the case of no decoherence effects,  the phase $\phi$
      accumulated along channel 0 is a deterministic variable. Otherwise,
      it is a stochastic variable described by the distribution given in Eq.~(\ref{distrib}).
      The electron-intensity measurements will show  single particle
      interference depending upon the phase accumulated and upon the noise effects.}
  \end{center}
\end{figure}

Here, we describe schematically  an electronic version of the ubiquitous
optical single-particle MZI~\cite{Born}
exploiting  electron transport in 1D
structures~\cite{Yang,Eugster,Bertoni,Zibold}.  The layout of the
single-particle interferometer is sketched in Fig.~\ref{fig1}. 

Specifically, we consider a system of two coupled electron channels with an electron injected
from the source $S$ and propagating along $x$ in the upper waveguide. 
Within the assumption of ballistic transport with no dephasing effect,  the carrier is described by a plane wave with wavevector $k$ along the $x$ direction, while  the qubit $|0\rangle$ ($|1\rangle$) is defined through the localization of the electron in the upper (lower) channel and can be expressed in terms of the localized states of a double-well potential.

The basic elements needed to perform the single-particle quantum operations are the phase shifter and the beam splitter. The former can be obtained by inserting a suitable potential barrier in one of the two channels, thus inducing a delay phase in the propagation of the wavefunction, as shown by a simple investigation of the electron transmission amplitude.
For a particle with kinetic energy $E$=$\hbar^2 k^2/2m$ and a potential barrier of height  $V_0$ and length $d$, the  transmission amplitude $t$ takes the form
\begin{equation}
  t=\frac{1}{\cos^2{k_1d} +\frac{1}{4}\left( \frac{k_1}{k}+\frac{k}{k_1}\right)^2 \sin^2{(k_1d)} } \exp(i\phi),
\end{equation}
where $k_1$=$\sqrt{2m (E-V_0)}/\hbar$, and
\begin{equation}
  \phi = -kd+\arctan{\left[\frac{1}{2}\left(\frac{k_1}{k} + \frac{k}{k_1}\right) \tan{k_1d}\right]} .
\end{equation}
In the limit of a small barrier, i.e. $V_0 \ll E$, $|t|^2$ is close to 1, that is the electron is entirely transmitted, while the delay phase $\phi$ reduces to $\phi=-(kd/2) (V_0/E)$. The matrix representation of the electronic phase shifter $R_{0(1)}$ on the one-qubit basis $\{|0\rangle, |1\rangle\}$ obtained by inserting a potential in the channel 0(1) will be given by
\begin{equation} \label{rotaz}
  R_{0}(\phi)= \left( \begin{array}{cc}
      e^{i\phi} & 0 \\
      0 & 1
    \end{array}\right) \qquad \textrm{and}\qquad  R_{1}(\phi)= \left( \begin{array}{cc}
      1& 0 \\
      0 & e^{i\phi}
    \end{array}\right).
\end{equation}

The electron beam splitter can be obtained by suitably tuning, in a coupling region along the propagation direction, the potential profile forming the two quantum channels~\cite{Bird}. Specifically, it has been shown that the carrier wavefunction in the transversal direction oscillates by lowering the potential barrier between the quantum wells defining the quantum channels in the transversal direction~\cite{Eugster,Bertoni}.
This process terminates when the carrier reaches the end of the coupling region and the electron wavefunction is separated into two components travelling along the upper and the lower channel.
The matrix representation of such one-qubit transformation reads
\begin{equation}\label{RX}
  R_y(\theta)= \left( \begin{array}{cc}
      \cos{\frac{\theta}{2}} & i\sin{\frac{\theta}{2}} \\
      i\sin{\frac{\theta}{2}} & \cos{\frac{\theta}{2}}
    \end{array}\right) ,
\end{equation}
where the phase $\theta$ depends upon the physical and geometrical
parameters of the coupling region and the electron energy~\cite{BertoniMod}.

\subsection{No decoherence}
Here, we investigate the behavior of the one-particle interferometer illustrated in Fig.~\ref{fig1},
when no source of electron decoherence  is considered. The electron arriving at 
the drains $D_{0}$ and $D_1$ is described by a pure single-particle state $|\Psi_{OUT}\rangle$.
This  is related to the input state $|\Psi_{INP}\rangle$=$|0\rangle$
by:
\begin{eqnarray} \label{layout}
  |\Psi_{OUT}\rangle &=& R_y\left(\frac{\pi}{2}\right) R_0(\phi) R_y\left(\frac{\pi}{2}\right)|\Psi_{INP}\rangle  \nonumber \\
  &=& i e^{i\frac{\phi}{2}}\left (\sin{\frac{\phi}{2}} |0\rangle +  \cos{\frac{\phi}{2}} |1\rangle\right).
\end{eqnarray}
where  $R_y\left(\frac{\pi}{2}\right) R_0(\phi) R_y\left(\frac{\pi}{2}\right)$ describes the network of quantum gates implementing the single-electron MZI.
In the measurements performed at the output drain $D_{0(1)}$, the detector response function is proportional to the square modulus of the 0(1) wavefunction component, namely
\begin{equation}
|\sin{\frac{\phi}{2}}|^2 (|\cos{\frac{\phi}{2}}|^2).
\end{equation}
Such oscillations depending upon $\phi$ in the electron intensity (or in the coherent component of the current) are a manifestation of the single-particle interference.  The visibility of interference patterns, given by $\nu$=$\left(I{_\textrm{max}}-I_{\textrm{min}} \right)/\left(I_{\textrm{max}}+I_{\textrm{min}} \right)$, where $I_{\textrm{max(min)}}$ indicates the maximum (minimum) signal intensity  revealed by a charge detector in the drain $D_{0(1)}$, is equal to the optimal value of 1.

\subsection{Decoherence}
The effects due to the  loss of electronic coherence on the functioning of the  single-electron MZI sketched in Fig.~\ref{fig1} are now examined.

In our scheme, the classical noise is described in terms of the fluctuations  of the potential profile along the propagation direction of the carriers, while along $y$  the double-well potential is assumed to be immune from noisy effects. This means that no undesired channel mixing due to environment action can occur, only those noise mechanisms leading to a randomization of the phase of the carriers are considered. 

As shown above, the  potential barrier  inserted in  channel 0 induces a delay phase  linearly depending upon the width and the height of the barrier itself. In close analogy with the procedure used in Ref.~\cite{Ady}, the potential barrier  parameters are supposed to fluctuate according to a stochastic behavior as a consequence of the noise.
This also implies the randomness of the delay phase $\phi$.
For the sake of simplicity, the latter is assumed to be described by a flat distribution
\begin{equation} \label{distrib}
  P(\phi)=\left\{ \begin{array} {l} 1/\Delta\phi \qquad \textrm{for} \qquad |\phi-\phi_0| \le \Delta\phi \\ 0  \qquad \textrm{otherwise}
                  \end{array} \right .,
\end{equation}
where $\phi_0$ is the  delay phase value in the absence of noise  and $\Delta \phi$ is a measure of the decoherence effects induced by the environment, that is of  the ``noise intensity''. In fact, when  $\Delta\phi$ goes to zero the noise effect vanishes, while for  $\Delta\phi=2\pi$ the  maximum dephasing is obtained. The choice of a flat probability distribution allows one to solve analytically  the system, and therefore, to provide an initial guideline for the analysis of decoherence phenomena in more general cases.

\begin{figure}[htpb]
  \begin{center}
    \includegraphics*[ width=\linewidth]{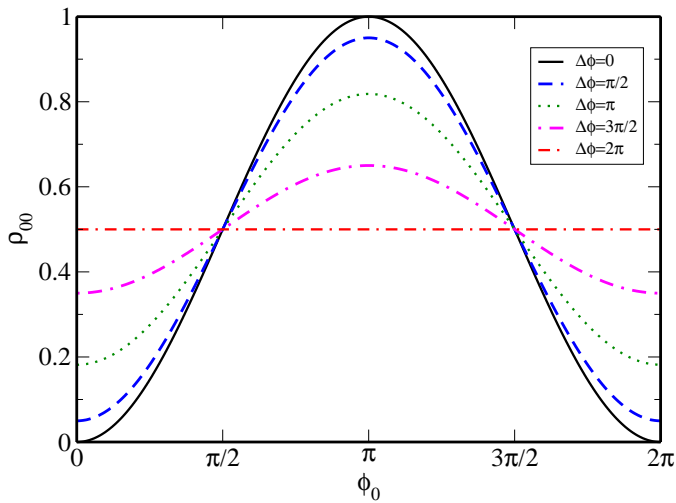}
    \caption{\label{fig2} Single-particle probability of finding the
      electron of the MZI in the drain $D_0$ as a
      function of the phase $\phi_0$ for different values of  $\Delta\phi$,
      ranging from 0, no decoherence,
      to $2\pi$, maximum decoherence.}
  \end{center}
\end{figure}

In strict analogy with the random field approaches introducing classical noise in quantum system~\cite{Corato,Bene,Amir}, the quantum state describing the electron at the output drains will be given by the statistical mixture
\begin{equation} \label{aver}
  \overline{\rho}=\frac{1}{\Delta\phi}\int_{\phi_0-\frac{\Delta\phi}{2}}^{\phi_0+\frac{\Delta\phi}{2}} d\phi |\Psi_{OUT}(\phi)\rangle \langle \Psi_{OUT}(\phi)|,
\end{equation}
obtained from  the average over a number of density matrices  $|\Psi_{OUT}(\phi)\rangle \langle \Psi_{OUT}(\phi)|$ each one corresponding to a specific delay phase $\phi$. 
By substituting  Eq.~(\ref{layout}) in (\ref{aver})  and after a straightforward calculation, we find  that $\overline{\rho}$ takes the form:
\begin{equation}
  \overline{\rho}= \frac{1}{2}\left( \begin{array}{cc}
      1- \frac{\sin{\Delta\phi/2}}{\Delta\phi/2} \cos{\phi_0} & \frac{\sin{\Delta\phi/2}}{\Delta\phi/2}  \sin{\phi_0}\\
       \frac{\sin{\Delta\phi/2}}{\Delta\phi/2}  \sin{\phi_0}     &  1+\frac{\sin{\Delta\phi/2}}{\Delta\phi/2} \cos{\phi_0} 
    \end{array}\right).
\end{equation}
The diagonal elements  of  $\overline{\rho}$ describe the probability of finding one electron at the output drains $D_0$ and $D_1$, that, in turn, correspond to the signal intensity revealed at the detectors.
Such an intensity   depends  upon both the delay phase $\phi_0$ and the noise parameter $\Delta\phi$, as shown in Fig.~\ref{fig2}. 
Once $\overline{\rho}$ has been obtained, the visibility of the single-electron can be evaluated. Its expression reads:
\begin{equation} \label{visi}
  \nu=\frac{\sin{\Delta\phi/2}}{\Delta\phi/2},
  \end{equation}
that is a decreasing function of $\Delta\phi$ in the interval $[0,2 \pi]$.
When no noise  is considered,  namely $\Delta\phi$=0, the results of the previous subsection are found, that is,  $\rho$ reduces to $|\Psi_{OUT}(\phi_0)\rangle \langle \Psi_{OUT}(\phi_0)|$ and $\nu$=1.
By increasing $\Delta\phi$, the effects of the environmental noise become stronger and the visibility of single-particle interference patterns decreases, as shown in Fig.~\ref{fig2}. When $\Delta\phi$=$2\pi$, the off-diagonal elements of $\overline{\rho}$ vanish while the diagonal ones tend to 1/2, and $\nu$ goes to zero. In this case, the noise destroys completely the quantum interference between the components of the output single-particle wavepackets.

These results  suggest that a relation between local single-particle quantities ascribed to wave-particle duality and  bipartite non-local properties (loss of coherence) must hold in the system under investigation,  in agreement with the well-known complementarity  relation  introduced for  pure two- and multi-qubit  states~\cite{Jag1,Jag2,Peng2}.
In the case of mixed single-particle states, such relation becomes
\begin{equation}  \label{p2c2}
  P^2 +\nu^2+C_G^2=1,
\end{equation}
where   $P$ indicates the predictability of the system: in our case a measure of the \emph{a priori} knowledge of whether the electron in the interferometer  is in the state $|0\rangle$ or $|1\rangle$. While the visibility $\nu$ is related to the wavelike nature of the system, $P$ quantifies the particle-like behavior. 
$C_G$ is  the  generalized concurrence which provides a measure of the  decoherence undergone by the electron as a consequence of its coupling with the environment, and it is  given by $C_G=\sqrt{2(1-\textrm{Tr}\overline{\rho}^2)}$.

Let us verify that the relation~(\ref{p2c2}) is satisfied in  the setup under investigation.
In analogy  with the approach used in Ref.~\cite{Peng}, the predictability can be evaluated as 
$P$=$\left|\langle \Theta |\sigma_z|\Theta\rangle \right|$,
where $\sigma_z$=$\left (\begin{array}{cc} 1 & 0\\ 0 & -1 \end{array}\right)$
and $|\Theta\rangle$=$\frac{|0\rangle +i |1\rangle } {\sqrt{2}}$  is the state describing the electron after crossing the first beam splitter. Our evaluation  yields $P$=0. On the other hand, the generalized concurrence results to be
\begin{equation}
  C_G=\sqrt{1-\left(\frac{\sin{\Delta\phi/2}}{\Delta\phi/2}\right)^2}.
\end{equation}
By taking into account  Eq.~(\ref{visi}), the complementarity relation~(\ref{p2c2}) holds for  the single-electron mixed state of our setup, as expected.
This implies that the loss of coherence of the system can be estimated by means of the single-particle detector response.

\section{Two-particle interferometer with no noise  }
\label{TPI}
In this section, we  first describe a  two-electron interferometer given by the Coulomb  coupling of two single-electron MZIÕs.  Then, we focus on  the close relation between interference effects and two-particle entanglement  in absence of environmental noise.

\begin{figure}[htpb]
  \begin{center}
    \includegraphics*[width=\linewidth]{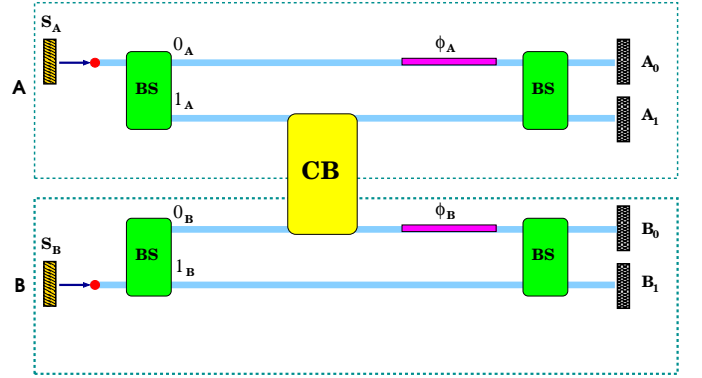}
    \caption{\label{fig3} Sketch of the electron waveguide network
      used to implement the 2EMZI. The two electrons are injected from the
      sources $S_A$ and $S_B$, and each one propagates
      in one couple of channels, i.e. a single MZI, involving two beam
      splitters and a delaying barrier potential inducing a change
     of  phase of $\phi_A$, and $\phi_B$ for the upper and lower
      interferometer, respectively.  In  case no decoherence effects are considered,  the phase $\phi_{A(B)}$
      accumulated along channel 0 of the $A(B)$ MZI  is a deterministic variable. Otherwise,
      it is a stochastic variable described by a flat distribution.
      The coupling block (CB)  correlates  the
      qubit states of the two single-particle MZI's by means of the Coulomb interaction between the
      electron wavefunctions propagating along the inner channels. Finally, the
      carrier injected by the source $S_A$ ($S_B$) is collected by the
      drains $\{A_0,A_1\}$ ($\{B_0,B_1\}$). }
  \end{center}
\end{figure}

A diagram of the two-electron MZI (2EMZI) is displayed in
Fig.~\ref{fig3}.  Two  charge carriers are emitted by
the sources $S_A$ and $S_B$ into two single-particle interferometers,
namely $A$ and $B$. Beyond the one-qubit
gates, the system involves a conditional phase gate consisting of a
coupling block which correlates, via Coulomb interaction, the qubit
states of the electrons propagating in the subsystems $A$ and $B$.
Coulomb interaction between electrons has already been proposed as
mechanism to entangle waveguide-based qubits and therefore to
implement a two-qubit gate~\cite{Bertoni,Rei,Akg}.

The coupling block sketched in Fig.~\ref{fig3} consists of a section
of length $l$ where the inner waveguides, namely $1_A$ and $0_B$, are
brought close to each other in order to ``activate'' the Coulomb
repulsion between the electrons travelling in those channels.
Without loss of generality, we adopt a model of  Coulomb gate already introduced in
Ref.~\cite{Akg}.  The Coulomb  energy potential between the two
particles in the inner channels has been taken proportional to the
inverse of the distance $D$ within the interaction window, i.e.
$V_{CO}=e^2/(4\pi\epsilon D)$, and zero elsewhere.  In such a scheme,
each of the two particles crossing the coupling block feels a
rectangular barrier potential  $V_{CO}$ high and $l$  wide.  Thus the
phase change $\gamma$ in the network state $|0_B 1_A \rangle$ related
to electron travelling through the interaction window can be analytically  evaluated
by means of the time-independent approach already used in the
Sec.~\ref{SPI} to analyze the electron phase shifter. Within the
approximation of electron energies much larger than $V_{CO}$ in order
to prevent reflections, the transmission probability
results about 1, while the phase change is given by $\gamma=- \sqrt{\frac{2m}{E}} \frac{e^2}
{8 \pi \hbar \epsilon D}$ and, as expected, depends both
upon the physical parameters of the carriers and the geometry of the
coupling block. In turn, these affect the amount of quantum
correlations created between the two
particles. In the following,
we will consider values of $\gamma$ ranging from 0 to $2\pi$.

The matrix representation of the conditional phase gate 
discussed above in the two-qubit base $\{|0_A 0_B\rangle$, $|0_A
1_B\rangle$, $|1_A 0_B\rangle$, $| 1_A1_B\rangle\}$ is
\begin{equation}
  T(\gamma)= \left( \begin{array}{cccc}
      1 & 0 & 0 & 0 \\
      0 & 1 & 0 & 0  \\
      0 & 0 & e^{i\gamma} & 0 \\
      0 & 0 & 0 & 1
    \end{array}\right) .
\end{equation}

Once crossed the coupling block, the electrons propagating in the 1D
waveguides interact with the two potential barriers placed in the
$0_A$ and $0_B$ channels, inducing a phase change of $\phi_A$ and
$\phi_B$, respectively. Then, the particles interfere again 
 and finally are collected by
the detectors at the output drains $\{A_0,A_1,B_0,B_1\}$. Thus, the
global logical transformation on the two-particle input state
$|\Phi_{IN} \rangle$=$|0_{A} 1_{B}\rangle$ can be written as:
\begin{equation}
  R_y^A\left(\frac{\pi}{2}\right)R_y^B\left(\frac{\pi}{2}\right)
  R_{0}^B(\phi_B) R_{0}^A(\phi_A) T^{AB}(\gamma) R_y^A\left(\frac{\pi}{2}\right)R_y^B\left(\frac{\pi}{2}\right),
\end{equation}
where the superscripts of the quantum gates indicate which subsystem they act on. After a straightforward calculation, we find that the output state $|\Phi_{OUT} \rangle$ takes the form
\begin{eqnarray} \label{OUT}
|\Phi_{OUT} \rangle 
&=& \sum_{X,Y=0}^1 \alpha_{XY}(\gamma) | X_A Y_B\rangle
= \frac{1}{2}e^{i\left(\frac{\phi_B}{2}+\phi_A\right)} \times  \\
&& \bigg[i\!\left(\cos{\frac{\phi_B}{2}} - e^{i\left(\frac{\gamma}{2}-\phi_A\right)} \cos{\frac{\phi_B+\gamma}{2}}\right) |0_{A} 0_{B}\rangle \nonumber \\
&& +i\left(e^{i\left(\frac{\gamma}{2}-\phi_A\right)} \sin{\frac{\phi_B+\gamma}{2}}-\sin{\frac{\phi_B}{2}}\right)|0_{A} 1_{B}\rangle \nonumber \\
&& -\left(\cos{\frac{\phi_B}{2}}+e^{i\left(\frac{\gamma}{2}-\phi_A\right)}\cos{\frac{\phi_B +\gamma}{2}}\right) |1_{A} 0_{B}\rangle \nonumber \\
&& +\!\!\left(e^{i\left(\frac{\gamma}{2}-\phi_A\right)}\sin{\frac{\phi_B +\gamma}{2}}+\sin{\frac{\phi_B}{2}}\right)|1_{A} 1_{B}\rangle\bigg]. \nonumber
\end{eqnarray}
Such an expression describes the two electrons at the output drains in a non-separable state where the amount of quantum correlations between the qubits depends upon the phase $\gamma$.

\subsection{Entanglement}
Here the carrier-carrier entanglement is evaluated in terms of the Wotters concurrence~\cite{Wooters} obtained from the two-particle density matrix $\rho^{AB}$ as 
\begin{equation}
C=\max{\{0, \sqrt{\lambda_1}-\sqrt{\lambda_2}-\sqrt{\lambda_3}-\sqrt{\lambda_4}\}},
\end{equation}
where $\lambda_i$ are the eigenvalues of the matrix $\zeta$
\begin{equation}  \label{autova}
  \zeta=\sqrt{\rho^{AB}} (\sigma_y^A \otimes \sigma_y^B) {\rho^{AB}}^{\ast} (\sigma_y^A \otimes \sigma_y^B) \sqrt{\rho^{AB}}
\end{equation}
arranged in decreasing order. 
Here, ${\rho^{AB}}^{\ast}$ denotes the complex conjugation of $\rho^{AB}$ in the standard basis, and $\sigma_y^{A(B)}$ is the well-known Pauli matrix acting on the qubit state of the subsystem $A(B)$. 
$C$ ranges from 0, for a disentangled state, to 1, for a maximally correlated state. 
For a pure two-qubit state, the concurrence  corresponds to  the generalized concurrence $C_G$ defined in the previous section. Indeed, the qubit-qubit entanglement can also be interpreted as the  loss of decoherence undergone by a qubit as a consequence of its interaction with the other qubit considered as the environment.

For   the output state $\rho^{AB}$ =$|\Phi_{OUT} \rangle\langle \Phi_{OUT}|$ describing the electrons at the drains, $C$ is given by
\begin{equation} \label{cocu}
  C=\sin{\frac{\gamma}{2}}.
\end{equation}
When $\gamma=0$, the output state, does not exhibit quantum correlations.
On the other hand, for $\gamma$=$\pi$  a maximally entangled state is obtained.
As a matter of fact, if now the two electrons at drains are processed by the network of one-qubit gates
\begin{displaymath}
  R_y^B\left(\frac{\pi}{2}\right) 
  R_{0}^B(-\phi_B) R_{0}^A(-\phi_A) R_y^B\left(-\frac{\pi}{2}\right)R_y^A\left(-\frac{\pi}{2}\right)
\end{displaymath}
the Bell state $i/\sqrt{2}\left( |0_A 0_B\rangle +|1_A 1_B\rangle \right)$ is produced.

\subsection{Interference} \label{SSint}
Let us focus our attention on the single-particle quantum interference described by Eq.~(\ref{OUT}).  To this aim, we need to estimate the one-particle reduced density matrices $\rho^{A(B)}$=$\textrm{Tr}_{B(A)} \rho^{AB} $.
After straightforward calculations, we find:
\begin{equation}\label{rhoa}
  \rho^A= \frac{1}{2}\left( \begin{array}{cc}
      1- F(\phi_A,-\gamma)  & G(\phi_A,-\gamma)  \\
       G(\phi_A,-\gamma)   &   1+ F(\phi_A,-\gamma)
    \end{array}\right),
\end{equation}
and
\begin{equation} \label{rhob} \rho^B= \frac{1}{2}\left( \begin{array}{cc}
      1+F(\phi_B,\gamma)  & -G(\phi_B,\gamma)  \\
       -G(\phi_B,\gamma)   &   1-F(\phi_B,\gamma)
    \end{array}\right),
\end{equation}
where 
\begin{eqnarray}
F(\phi,\gamma)&=&\frac{1}{2} \left[ \cos{\phi}+\cos{(\phi+\gamma}) \right], \nonumber \\
G(\phi,\gamma)&=&\frac{1}{2} \left[  \sin{\phi}+\sin{(\phi+\gamma}) \right].
\end{eqnarray}
For both  interferometers, the visibility of the 
single-particle interference fringes is given by 
\begin{equation} \label{visiab}
  \nu_{A(B)}=\left|\cos{\frac{\gamma}{2}}\right|.
  \end{equation}
When no qubit-qubit entanglement is
created, namely $\gamma$=0,  the two interferometers are completely
uncorrelated and the electron detection probabilities are described by sinusoidal functions $\sin^2{\frac{\phi_{A(B)}}{2}}$ and
$\cos^2{\frac{\phi_{A(B)}}{2}}$ of the local phase $\phi_{A(B)}$. In this case, the visibility takes its maximal value 1.
The more correlated are the particles  (i.e. the greater  is the loss of coherence undergone by an electron as a consequence of
the electron-electron interaction) the smaller  is the visibility of single-electron
interference.  When  entanglement reaches its maximum value, quantum interference between
the components of the output single-particle wavepackets is completely
destroyed. This is in agreement with the results of Sec.~\ref{SPI} for the single-electron MZI subject
to environmental noise.

In agreement with the theoretical predictions~\cite{Jab2,Suz}, the complementarity relation
\begin{equation}  \label{C2EMZI}
C^2+\nu_{A(B)}^2+P^2_{A(B)}=1,
\end{equation}
holds in the 2EMZI. 
The predictability $P_{A(B)}$=$\left|\langle \Xi |\sigma_z^{A(B)}|\Xi\rangle\right|$
($|\Xi\rangle$ describes the two electrons after the
coupling block) is zero like the case of single-electron MZI.
 Therefore, here the qubit-qubit entanglement
can  simply be related to the single-particle visibility as $C$=$\sqrt{1-\nu^2}$.
The expression gives a useful tool to quantify  the amount of quantum correlations
and therefore  the measure of the loss of single-particle coherence due to electron-electron
interaction. Such a measurement relies on  single-particle detectors response in an experimental setup  of 2EMZI where decoherence effects
are negligible.    

\begin{figure}[htpb]
  \begin{center}
    \includegraphics*[width=\linewidth]{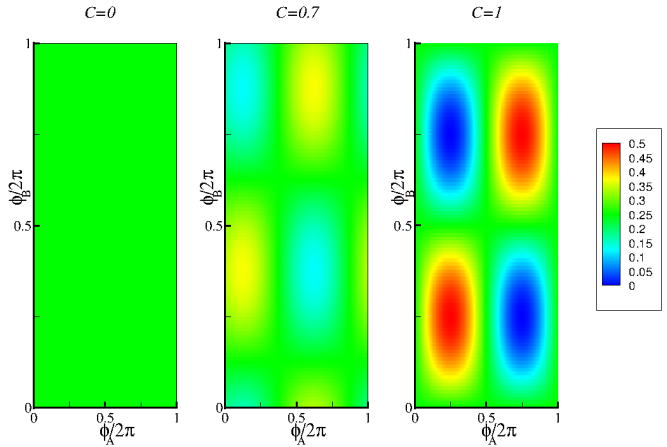}
    \caption{\label{fig4} The ``corrected'' joint probability $P_{00}$
      of finding two electrons at drains $A_0$ and $B_0$ as a function
      of $\phi_A$ and $\phi_B$ for three different values of
      $C$, namely 0, 0.7, and $1$. The visibility of
      two-particle fringes increases with $C$.
      Note that the tailoring of $C$ affects not only the magnitude
      of $P_{00}$ but also its dependence on the two phases.}
  \end{center}
\end{figure}

A complementary relation of the type given in Eq.~(\ref{C2EMZI}) can also
be established between one-particle and two-particle interference, as
argued by Jager~\emph{et al.}~\cite{Jag1,Jag2}.  Two-particle effects
are related to the joint probability of detecting one electron at
$A_X$ and the other one at $B_Y$ in a measure
performed by detectors at the output drains. However,
such a detection probability, given by the square modulus of the coefficient
$\alpha_{XY}$,  can exhibit non \emph{genuine} two-particle interference related 
to the single-particle probabilities of finding the electron of the
interferometer $A$ in the channel $A_X$ and the other electron in the
channel $B_Y$. In order to evaluate \emph{genuine} two-particle effects, 
some authors~\cite{Jag1,Jag2} proposed
to use a ``corrected'' joint two-particle probability $P_{XY}$ given
by
\begin{equation} \label{twoP}
  P_{XY}= |\alpha_{XY}|^2-\rho^A_{XX}\rho^B_{YY}+\frac{1}{4}.
\end{equation}
In Fig.~\ref{fig4} we have reported $ P_{00}$ as a function of the
phases $\phi_A$ and $\phi_B$. While for zero entanglement, such a
probability assumes  the constant value $1/4$ for all the phases values, two-particle fringes
become more and  more visible as the degree of entanglement increases. 
In fact, $ P_{00}$ can be used to quantify the two-particle visibility
$\nu_{12}$:
\begin{equation}
  \nu_{12}=\frac{[P_{00}]_{\textrm{max}}      - [P_{00}]_{\textrm{min}}}{[P_{00}]_{\textrm{max}}     + [P_{00}]_{\textrm{min}}}=\sin^2{\frac{\gamma}{2}}=C^2,
\end{equation}
which results to be the square of the concurrence, in agreement with the results presented in the literature
showing the close relation between entanglement and two-particle
visibility~\cite{Jab2}.  As expected,  the two-particle visibility is zero for
$\gamma$=0, i.e.  the two-particle interference effects are due only
to single-particle fringes. On the other hand, $\nu_{12}$ reaches the maximum
value for $\gamma$=$\pi$, when the electron-electron correlation gives
rise to genuine two-particle fringes.  In
agreement with the well-known complementarity
inequality~\cite{Jag1,Jag2},  an increase of $\nu_{12}$ corresponds to a decrease of the visibility of
the one-particle fringes, and vice versa:
\begin{equation}
  \nu_{A(B)}^2+ \nu_{12}^2=\cos^2{\frac{\gamma}{2}}+\sin^4{\frac{\gamma}{2}}\le 1.
\end{equation}
$\nu_{12}$ could be also used, at least  in principle, to quantify the
degree of the qubit-qubit entanglement.  Nevertheless, in an
experimental implementation of our setup  the estimation of the
visibility of two-particle interference would certainly be more
complex than the one of $\nu$, since it requires the simultaneous 
detection of two electrons in two different detectors.

\section{Effects of noise on a two-particle interferometer }
\label{TPI2}
Here, we analyze how the environmental noise affects the interference and entanglement of the setup sketched in Fig.~\ref{fig3}.

In agreement with the approach used in Sec.~\ref{SPI}, the noise is assumed to randomize the  phases $\phi_{A}$ and $\phi_{B}$ of the two carriers.  Such dephasing effects  are independent of each other since the fluctuations of the potential  barriers of the channels 0 of the two MZIs are assumed to be due to different noise sources. 
$\phi_{A(B)}$ is described by the flat distribution $P(\phi_{A(B)})$ of Eq.~(\ref{distrib}) with parameters ${\phi_0}_{A(B)}$ and $\Delta\phi_{A(B)}$.  The two particles arriving at the drains $\{A_0,A_1,B_0,B_1\}$ will be in the mixed state 
\begin{eqnarray} \label{douint}
 \overline{\rho^{AB}} &=& \frac{1}{\Delta\phi_A \Delta\phi_B} \int_{{\phi_0}_A-\frac{\Delta\phi_A}{2}}^{{\phi_0}_A+\frac{\Delta\phi_A}{2}} d\phi_A \times \\
 && \int_{{\phi_0}_B-\frac{\Delta\phi_B}{2}}^{{\phi_0}_B+\frac{\Delta\phi_B}{2}} d\phi_B  |\Phi_{OUT} (\phi_A, \phi_B )\rangle\langle \Phi_{OUT} (\phi_A, \phi_B )|. \nonumber
\end{eqnarray}
Its explicit evaluation is given in Appendix~\ref{App}. 
For the sake of simplicity, we will assume that the spectra of the environmental noises on the two MZIs, namely their dephasing, are identical even if they stem from different sources. Such an  assumption means that $\Delta\phi_{A}$=$\Delta\phi_{B}$=$\Delta\phi$.

\subsection{Entanglement}
\begin{figure}[htpb]
  \begin{center}
    \includegraphics*[ width=\linewidth]{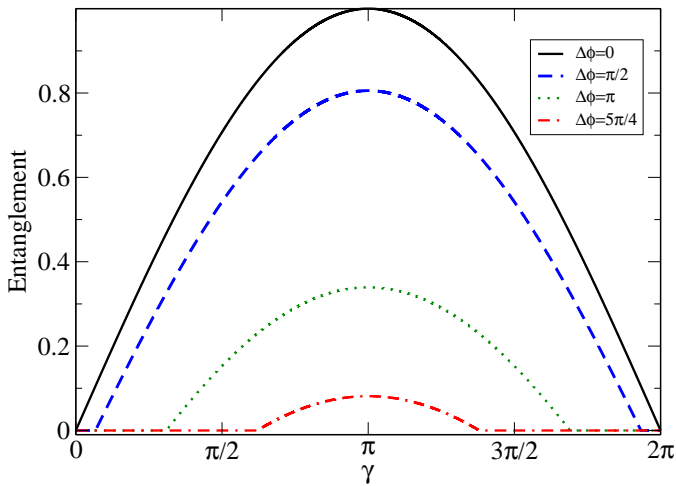}
    \caption{\label{newfig} Electron-electron entanglement as a function
    of the Coulomb-induced phase $\gamma$  evaluated for different values of  $\Delta\phi$ indicated in the legend.
    Note that for $\Delta\phi\ge \Delta\phi_c \approx1.33 \pi$ entanglement is identically zero for any value
    of $\gamma$.}
  \end{center}
\end{figure}

\begin{figure}
\begin{center}
    \includegraphics*[ width=\linewidth]{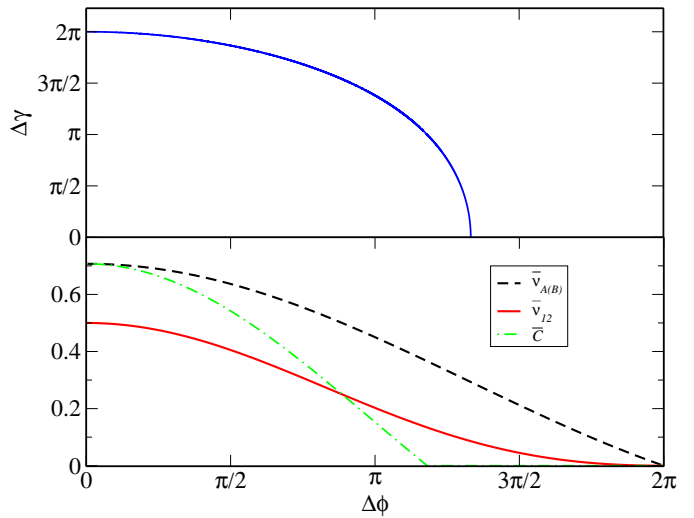}
    \caption{\label{graph} Top panel: The interval $\Delta\gamma$ over which entanglement takes non zero  values as a function of 
   $\Delta\phi$ . Bottom panel: single-particle visibility $\overline{\nu}_{A(B)}$ (dashed line), two-particle visibility $\overline{\nu}_{12}$
   (solid line), and entanglement $\overline{C}$ ( dash-dotted line) as a function of  $\Delta\phi$ with $\gamma$=$\pi/2$.}
  \end{center}
\end{figure}

First, we estimate the qubit-qubit entanglement in terms of concurrence defined in the previous section. By inserting  Eqs.~(\ref{rhoabn}) in Eq.~(\ref{autova}),  we can compute the concurrence for the mixed state of Eq.~(\ref{douint})
\begin{equation} \nonumber
  \overline{C}=\max{\{0,g\}},
\end{equation}
where
\begin{equation}\label{frt}
  g= \frac{\sin{\Delta\phi/2}}{\Delta\phi/2}\sin{\frac{\gamma}{2}} -\frac{1}{2}\left[1-\left(\frac{\sin{\Delta\phi/2}}{\Delta\phi/2}\right)^2\right].
\end{equation}
In Fig.~\ref{newfig}, we have reported $\overline{C}$ as a function of $\gamma$ for different values of $\Delta\phi$. 
As expected, for zero dephasing  the concurrence exhibits a sinusoidal dependence against the phase $\gamma$. This allows us to quantify the qubit-qubit entanglement and to expose its reduction with the noise.  Indeed, in this case expression~(\ref{frt}) reduces to  Eq.~(\ref{cocu}) obtained  for the 2EMZI in absence of noise. When noise effects are taken into account, the  entanglement exhibits a peculiar behavior. Not only  the maximum amount of quantum correlations created between two qubits decreases as $\Delta\phi$ increases, but $\overline{C}$ vanishes even when  the Coulomb coupling between the two MZIs is different from zero. Specifically,  the $\Delta \gamma$ phase interval  over which electron-electron entanglement takes non zero values is related to $\Delta\phi$ by the expression
\begin{equation} \label{frt2}
  \Delta \gamma=2\pi-4\arcsin{\left(\frac{1-\left(\frac{\sin{\Delta\phi/2}}{\Delta\phi/2}\right)^2}{\frac{2\sin{\Delta\phi/2}}{\Delta\phi/2}}      \right)}.
\end{equation}
As shown in the top panel of  Fig.~\ref{graph}, $\Delta \gamma$ decreases
rapidly with noise intensity and goes to zero when  the carrier dephasing induced
by the environment reaches the critical value $\Delta\phi_c\approx1.33 \pi$,
where the latter value is obtained from the relation $\frac{\sin{\Delta\phi_c/2}}{\Delta\phi_c/2}$=$\sqrt{2}-1$.
When  $\Delta\phi = \Delta\phi_c$,  the degree of entanglement is identically zero
for any value of $\gamma$. In other terms, in our setup
there is no need to reach the maximum dephasing effects
in order to inhibit completely   the entanglement production.
These results  clearly indicate that the randomization of the phase carrier is very effective into destroying  the quantum correlation
created between the two carriers as a consequence of their mutual repulsion. In this view, the entangled
states  produced in our device thanks to Coulomb coupling are not  very robust
under decoherence phenomena.

\subsection{Interference}
Now, we examine the effects of noise on the visibility  of the fringes
of  one- and two-particle  interference in the 2EMZI.  Following the approach
adopted in the previous section, the former can be evaluated from the single-particle
reduced density matrix  $\overline{\rho^{A(B)}}$ describing  the electron of subsystem $A(B)$.
This can simply be expressed in terms of $\rho^{A(B)}$ as
\begin{equation}
\overline{\rho^{A(B)}}_{ij}=\frac{\sin{\Delta\phi/2}}{\Delta\phi/2} \rho^{A(B)}_{ij} +\delta_{ij}\frac{1}{2}\left(1-\frac{\sin{\Delta\phi/2}}{\Delta\phi/2}\right),
\end{equation}
where $\delta_{ij}$ is the Kronecker  delta.
After a straightforward calculation, we get  the visibility of the one-particle interference:
\begin{equation} \label{visioise}
\overline{\nu}_{A(B)}=\nu_{A(B)}\frac{\sin{\Delta\phi/2}}{\Delta\phi/2}=\left|\cos{\frac{\gamma}{2}}\right|\frac{\sin{\Delta\phi/2}}{\Delta\phi/2}.
\end{equation}
$\overline{\nu}_{A(B)}$ is  given by the product of the  visibility $\nu_{A(B)}$, found  for the system in absence of noise (see
Eq.~(\ref{visiab}) )  and related to the Coulomb interaction by means of the phase $\gamma$,
and of the factor $\frac{\sin{\Delta\phi/2}}{\Delta\phi/2}$, which represents  the visibility of
a MZI, not coupled to other systems in the  presence of environmental noise inducing  phase randomization
of the carriers, as shown in Eq.~(\ref{visi}). Eq.~(\ref{visioise})  is in agreement with the fact
that, for the setup under investigation, the two decoherence channels
of  the  single-electron wavefunctions, one due to electron-electron interaction and the other one due
to the environmental noise, are uncorrelated. Therefore,  the effects of these two mechanisms on 
the single-electron visibility  are independent of each other.

By suitably revising Eq.~(\ref{p2c2}), the complementarity relation 
between single-particle and non-local quantities is still satisfied.
As stated above, now two independent  decoherence channels
are active and therefore the qubit-qubit entanglement alone cannot be used
to quantify  the decoherence undergone by a single qubit.
The latter has to be evaluated in terms of the generalized concurrence $\overline{C_G}$,
defined in Sec.~\ref{SPI}, which allows one to  estimate  the total loss of coherence
 of the single-electron wavefunctions. For the 2EMZI subject to the environmental noise
 $\overline{C_G}^{A(B)}$ takes the form
\begin{equation} \label{gcn}
\overline{C_G}^{A(B)}=\sqrt{1-\left(\frac{\sin{\Delta\phi/2}}{\Delta\phi/2}\right)^2 \cos^2{\frac{\gamma}{2}}}.
\end{equation}
In comparison with the behavior exhibited by the two-electron entanglement, we note that
for a given value of $\gamma$ the one-particle decoherence is maximum ($C_G$=1)
 only when noise effects are maximum ($\Delta \phi =2\pi$).
By taking into account that predictability vanishes,  Eqs.~(\ref{visioise}) and (\ref{gcn})
imply that the complementarity relation between visibility and loss of single-particle coherence still holds
for the mixed state produced in our  2EMZI.

Finally, we investigate the two-particle interference. In order to discriminate \emph{genuine} two-electron
effects from the ones deriving from single-particle detection probabilities when the noise effects
are present, we use a suitable  revised form of  Eq.~(\ref{twoP}):
\begin{equation} 
  \overline{P}_{XY}= \overline{\rho^{AB}}_{XYXY}-\overline{\rho^A_{XX}}\quad\overline{\rho^B_{YY}}+\frac{1}{4},
\end{equation}
where  $\overline{\rho^{AB}}_{XYXY}$ has been introduced  to describe
the joint probability of detecting one electron at drain $A_X$ and the other one
at $B_Y$.  As expected,  the \emph{genuine} two-particle interference effects,
described by   $\overline{P}_{XY}$, are affected  by noise. Two-particle  fringes 
are less visible for  higher disorder, as clearly indicated by the two-particle
visibility 
\begin{equation} \label{ert}
  \overline{\nu}_{12}=\left(\frac{\sin{\Delta\phi/2}}{\Delta\phi/2}\right)^2 \sin^2{\frac{\gamma}{2}}
\end{equation}
which goes to zero for $\Delta\phi$=$2\pi$.
The  complementarity inequality
relating $\overline{\nu}$ to $\overline{\nu}_{12}$ is still valid:
\begin{equation}
  \overline{\nu}_{A(B)}^2+ \overline{\nu}_{12}^2=\left(\frac{\sin{\Delta\phi/2}}{\Delta\phi/2}\right)^2\cos^2{\frac{\gamma}{2}}+
  \left(\frac{\sin{\Delta\phi/2}}{\Delta\phi/2}\right)^4\sin^4{\frac{\gamma}{2}} \le 1.
\end{equation}

Unlike the results   found
in Sec.~\ref{TPI},  here $\overline{\nu}_{12}$ cannot be used as an estimator of 
electron-electron entanglement. Indeed, two-particle interference can be detected at the output drains even
if  the degree of quantum correlation between the carriers is zero as indicated in the bottom panel of
Fig.~\ref{graph}. Nevertheless, we stress that in an experimental implementation 
of our setup the entanglement can still be estimated, at least in principle, by means  of
single-particle and two-particle detector responses. As shown
in Eq.~(\ref{frt}), entanglement depends only upon the two  independent parameters
$\gamma$ and  $\Delta\phi$ which appear in the expressions~(\ref{visioise}) and (\ref{ert}) 
describing visibility of one- and two-particle interference, respectively. An estimate of $\overline{\nu}$
and $\overline{\nu}_{12}$ permits to evaluate $\gamma$ and  $\Delta\phi$, and, in turn,  the entanglement.

\section{Conclusions}\label{Conclu}
Beyond its specific role in the experimental validation of 
quantum mechanics, electron transport in 1D waveguides  constitues an 
important resource  both for the realization of electronic interferometers~\cite{Tilke,Hei,Neder} and 
for  the design of  next-generation electron devices capable of processing quantum information~\cite{Bertoni,bus,bus2}{.
The main threat to these implementations is represented by dephasing
mechanisms which strongly affect the coherence of the quantum systems.

In this work, we have analytically 
investigated the interference phenomena and the appearance of  non-local effects
in  one- and two-electron interferometers realized by means of 1D channels and  subject
to an environmental noise.   
The latter stems from the fluctuations of the confining potential
of the interferometer arms and leads to a random detune of the transmission phase of the carriers.
As shown in the literature~\cite{Furlan,John},
nanometric few-electron systems are very sensitive to the random nature of intrinsic impurities and to the effects of quasi-static fluctuation of background charge.
Our model treats such noise classically by neglecting the building up of quantum entanglement between the above background and our system, namely the charge carriers along the waveguides. Furthermore, a single noise source is included in our calculations.
This has to be considered as a first step towards the fully quantitative modeling of a real device whose functioning relies on the coherence of the quantum system, 
with the inclusion of all the relevant sources of noise as, for example, amplifier noise, gate noise or elastic phonon scattering.


In order to validate our approach, first we  have studied  the dephasing  in an electronic  version of a single-particle MZI,  using  
only single-qubit logical operations. We  have found that the  single-electron visibility is related
to the loss of coherence induced by the environmental noise. Indeed, the greater are the noise effects, the smaller
is the visibility of  the single-electron  interference fringes. In the noisy MZI,  the well-know
complementarity relation~\cite{Jag1,Jag2} linking single-particle quantities to non-local quantities, such as the loss of coherence,
is  thus satisfied. This proves the feasibility of  our procedure to evaluate the features of dephasing
 in electron interference and entanglement in 1D channels.

In the second part of the paper, we have examined the two-electron interferometer
given by the coupling of two single-electron MZIs through  the  Coulomb
interaction. The latter was shown to yield   quantum correlations between the
carriers in other systems realized with 1D electron waveguides~\cite{Bertoni}. In the absence
of  environmental noise,  complementarity is found between 
the visibility of single-electron interference fringes and electron-electron entanglement, which
also represents an estimate of the loss of  quantum coherence of an electron due to its interaction
with the other carriers.  Quantum correlations stemming from the
Coulomb interaction between electrons affect also the  visibility of the \emph{genuine} two-particle interference fringes,
ax explained in Sec.~\ref{SSint}.
The latter turns out to increase with the square of the entanglement. 
These results suggest that the response of single-and two-particle detectors at the output drains could give a
quantitative estimation of the electron-electron entanglement  in an
experimental implementation of our system.

Finally, dephasing mechanisms have   been introduced in 
the  two-electron interferometer. We have found  that  the entangled states, 
obtained  thanks to the electron-electron interaction, 
are not very robust  with respect to  decoherence.  Indeed,
also for the case of optimal tuning of the geometrical parameters of the setup (leading to
maximum entanglement in absence of noise)
quantum correlations  between the carriers can be completely  destroyed by the environmental noise,
even when the latter does not yield maximum dephasing, thus partially preserving
single-electron interference.  This is a remarkable result, since it shows analytically that a small amount of noise
not only reduces entanglement but it completely suppresses quantum correlations unless the coupling
between the two subsystems is strong enough.

The visibility of the pattern of the single-particle interference results to be degraded by decoherence  due  
both to the mutual repulsion between the two carriers and to environmental noise effects. 
In particular, our results indicate that these mechanisms act as two independent and 
separate decoherence channels for one-electron states. As a consequence, 
the one-particle visibility  is   not  linked  by a complementarity  relation
to electron-electron entanglement but to   the total loss of coherence.
Noise plays also a key role in the analysis of \emph{genuine} two-particle interference fringes.
Their visibility, degrading with  the intensity of the noise, is somehow related to the one-electron visibility
by a suitable generalization of the complementarity inequality  holding for pure two-qubit states~\cite{Jag1,Jag2}.
Unlike what was found in absence of noise, the two-particle visibility 
 cannot be taken as an evaluator of the entanglement. Nevertheless,
in an experimental realization of the two-electron interferometer,
the estimate of the one- and two-particle visibility can be obtained  by means of the single- and 
cross-correlated detection of the carriers in two drains, respectively. Then, it  can be used to evaluate
the noise and Coulomb coupling parameters  and, from these, both entanglement
and decoherence. This shows that 1D channel architectures are a
powerful means to investigate entanglement and dephasing
in electron interference.

\appendix
\section{Two-particle density matrix of the noisy two-electron interferometer  }
\label{App}
Here we report the elements of the average density matrix describing two electrons
arriving at the  output drains in the noisy two-particle interferometer exploting
Coulomb interaction. The double integrals in the r.h.s of  Eq.~(\ref{douint}) yield
\onecolumn
\begin{displaymath}
\overline{\rho^{AB}}_{0000}=\frac{1}{4}\left[ 1+\frac{1}{2} f_B \bigg(\cos{{\phi_B}_0} +\cos{({\phi_B}_0+\gamma)}\bigg) -f_A \cos{\left({\phi_A}_0-\frac{\gamma}{2}\right)} \bigg(
 \cos{\frac{\gamma}{2}}+f_B \cos{\left({\phi_B}_0+\frac{\gamma}{2}\right)} \bigg)      \right] \
\end{displaymath}
\begin{displaymath}
\overline{\rho^{AB}}_{0101}=\frac{1}{4}\left[ 1-\frac{1}{2} f_B \bigg(\cos{{\phi_B}_0} +\cos{({\phi_B}_0+\gamma)}\bigg) -f_A \cos{\left({\phi_A}_0-\frac{\gamma}{2}\right)} \bigg(
 \cos{\frac{\gamma}{2}}-f_B \cos{\left({\phi_B}_0+\frac{\gamma}{2}\right)} \bigg)      \right] \
\end{displaymath}
\begin{displaymath}
\overline{\rho^{AB}}_{1010}=\frac{1}{4}\left[ 1+\frac{1}{2} f_B \bigg(\cos{{\phi_B}_0} +\cos{({\phi_B}_0+\gamma)}\bigg) +f_A \cos{\left({\phi_A}_0-\frac{\gamma}{2}\right)} \bigg(
 \cos{\frac{\gamma}{2}}+f_B \cos{\left({\phi_B}_0+\frac{\gamma}{2}\right)} \bigg)      \right] \
\end{displaymath}
\begin{displaymath}
\overline{\rho^{AB}}_{1111}=\frac{1}{4}\left[ 1-\frac{1}{2} f_B \bigg(\cos{{\phi_B}_0} +\cos{({\phi_B}_0+\gamma)}\bigg) +f_A \cos{\left({\phi_A}_0-\frac{\gamma}{2}\right)} \bigg(
 \cos{\frac{\gamma}{2}}-f_B \cos{\left({\phi_B}_0+\frac{\gamma}{2}\right)} \bigg)      \right] \
\end{displaymath}
\begin{eqnarray}
\!\!\!\!\!\!\!\!\!\!\!\!\!\!\overline{\rho^{AB}}_{0001}&=&\frac{1}{4}\bigg\{  f_B \left[f_A  \cos{\left({\phi_A}_0-\frac{\gamma}{2}\right)} \sin{\left({\phi_B}_0+\frac{\gamma}{2}\right)}- \frac{1}{2}\bigg(\sin{{\phi_B}_0} +\sin{({\phi_B}_0+\gamma)}\bigg)\right] +  \nonumber \\
& &+i f_A \sin{\frac{\gamma}{2}} \sin{\left({\phi_A}_0-\frac{\gamma}{2}\right)}\bigg\}  \nonumber
\end{eqnarray}
\begin{eqnarray}
\!\!\!\!\!\!\!\!\!\!\!\!\!\!\overline{\rho^{AB}}_{0010}&=&\frac{1}{4}\bigg\{  f_B \left[f_A  \sin\left({\phi_A}_0-\frac{\gamma}{2}\right)\cos{\left({\phi_B}_0+\frac{\gamma}{2}\right)}+\frac{i}{2}\bigg(\cos{({\phi_B}_0+\gamma)}-\cos{{\phi_B}_0} \bigg)\right] +  \nonumber \\
& &+f_A \cos{\frac{\gamma}{2}} \sin{\left({\phi_A}_0-\frac{\gamma}{2}\right)}\bigg\}  \nonumber
\end{eqnarray}
\begin{eqnarray}
\!\!\!\!\!\!\!\!\!\!\!\!\!\!\overline{\rho^{AB}}_{0011}&=&\frac{1}{4}\bigg\{ - f_B \left[f_A  \sin\left({\phi_A}_0-\frac{\gamma}{2}\right)\sin{\left({\phi_B}_0+\frac{\gamma}{2}\right)}+\frac{i}{2}\bigg(\sin{({\phi_B}_0+\gamma)}-\sin{{\phi_B}_0} \bigg)\right] +  \nonumber \\
& &+if_A \sin{\frac{\gamma}{2}} \cos{\left({\phi_A}_0-\frac{\gamma}{2}\right)}\bigg\}  \nonumber
\end{eqnarray}
\begin{eqnarray}
\!\!\!\!\!\!\!\!\!\!\!\!\!\!\overline{\rho^{AB}}_{0110}&=&\frac{1}{4}\bigg\{ - f_B \left[f_A  \sin\left({\phi_A}_0-\frac{\gamma}{2}\right)\sin{\left({\phi_B}_0+\frac{\gamma}{2}\right)}+\frac{i}{2}\bigg(\sin{({\phi_B}_0+\gamma)}-\sin{{\phi_B}_0} \bigg)\right] +  \nonumber \\
& &-if_A \sin{\frac{\gamma}{2}} \cos{\left({\phi_A}_0-\frac{\gamma}{2}\right)}\bigg\}  \nonumber
\end{eqnarray}
\begin{eqnarray}
\!\!\!\!\!\!\!\!\!\!\!\!\!\!\overline{\rho^{AB}}_{0111}&=&\frac{1}{4}\bigg\{ - f_B \left[f_A  \sin\left({\phi_A}_0-\frac{\gamma}{2}\right)\cos{\left({\phi_B}_0+\frac{\gamma}{2}\right)}+\frac{i}{2}\bigg(\cos{({\phi_B}_0+\gamma)}-\cos{{\phi_B}_0} \bigg)\right] +  \nonumber \\
& &+f_A \cos{\frac{\gamma}{2}} \sin{\left({\phi_A}_0-\frac{\gamma}{2}\right)}\bigg\}  \nonumber
\end{eqnarray}
\begin{eqnarray} \label{rhoabn}
\!\!\!\!\!\!\!\!\!\!\!\!\!\!\overline{\rho^{AB}}_{1011}&=&\frac{1}{4}\bigg\{ - f_B \left[f_A  \cos{\left({\phi_A}_0-\frac{\gamma}{2}\right)} \sin{\left({\phi_B}_0+\frac{\gamma}{2}\right)}+\frac{1}{2}\bigg(\sin{{\phi_B}_0} +\sin{({\phi_B}_0+\gamma)}\bigg)\right] +  \nonumber \\
& &-i f_A \sin{\frac{\gamma}{2}} \sin{\left({\phi_A}_0-\frac{\gamma}{2}\right)}\bigg\}, 
\end{eqnarray}
\twocolumn
where $f_{A(B)}$=$\frac{\sin{\Delta\phi_{A(B)}/2}}{\Delta\phi_{A(B)}/2}$ is the visibility of the single-particle interference
of the subsystem $A(B)$ when the two MZIs are not coupled.

\end{document}